\documentclass[
preprint,
nofootinbib,
amsmath,amssymb,
]{revtex4-2}
\usepackage{graphicx}
\usepackage{float}
\usepackage{multirow}
\def\be{\begin{equation}}
\def\ee{\end{equation}}
\def\bea{\begin{eqnarray}}
\def\eea{\end{eqnarray}}

\def\thatm{$\theta_{23}$}
\def\sinsqthatm{$\sin^{2}\theta_{23}$}

\def\threactor{$\theta_{13}$}
\def\sinsqthreactor{$\sin^{2}\theta_{13}$}
\def\dcp{$\delta_{\text{CP}}$}
\def\numu{$\nu_{\mu}$}
\def\numubar{$\overline{\nu}_{\mu}$}
\def\nue{$\nu_{e}$}
\def\nuebar{$\overline{\nu}_{e}$}

\begin{document}
\title{Long Baseline Neutrino Oscillation Results (T2K and NOvA)}

\author{ S. Cao (T2K collaboration) }
\email{cvson@ifirse.icise.vn}
\affiliation{The Institute For Interdisciplinary Research in Science and Education, ICISE \\
07 Science avenue, Ghenh Rang, Quy Nhon city, VN}

\date{\today}

\begin{abstract}
The well-established neutrino oscillation phenomenon, which confirms neutrinos have mass and the charged and neutral leptons are mixed, has been widely exploited to investigate the physics of this elusive particle. The complete description of neutrino oscillations, which are dictated by two mass-squared splittings, three mixing angles, and one Dirac CP-violation phase, however, has yet to be reached. The ongoing accelerator-based long-baseline neutrino experiments, T2K and NOvA, are critical in completing this picture. Using data and analyses accessible by summer 2023, T2K favors a leptonic CP violation at more than 90\% C.L. while NOvA shows no preference of this indication. Both experiments weakly opt for the \emph{normal} neutrino mass ordering and higher octant of the \thatm\ mixing angle.
\end{abstract}

\maketitle
\newpage

\section{Neutrino as a window on the Universe}
Neutrino is everywhere, dominated with the relic neutrino background ( $\approx 330\ \nu$ per $cm^3$). Other sources include the neutrinos from the sun and other astrophysical sources, from the reactors, from the Earth's atmosphere (\emph{due to interaction with cosmic rays}), and from the Earth's interior (\emph{or geo-neutrino due to decay of radionuclide}). The elusive neutrinos, the most abundant massive particles spanning in 24-order of energy magnitude~\cite{Vitagliano:2019yzm}, play essential role in the sensitivity, intensity, and energy frontiers. Since the first hypothesis of neutrino existence by W. Pauli in 1930 and a long-wait  experimental discovery in 1956, we have identified three types (\emph{or flavors}) of neutrinos $(\nu_e, \nu_{\mu}, \nu_{\tau})$ which are observed along with three corresponding leptons $(e,\ \mu,\ \tau)$. One of the most remarkable discoveries in the last 25 years is that the associated flavor of neutrino can be changed periodically as neutrino propagates. This phenomenon, known as neutrino oscillation, implies that neutrinos have mass and the charged and neutral leptons are mixed. The impacts of this discovery to the particle physics, astrophysics, and cosmology are enormous, details can be found elsewhere~\cite{ParticleDataGroup:2022pth}. The renormalizeable Standard Model, which predicts the zero mass of Dirac neutrino\footnote{Majorana mass term built with left-handed neutrino state is allowable}, must be extended. The extraordinary smallness of neutrino masses, the difference in the quark and lepton mixing patterns, and the non-hierarchical spectrum of neutrino mass may hint at the new physics. Massive neutrinos affect the evolution of the Universe and must be taken into account in analyzing the cosmological data. The high precision era of the cosmology, in turn, helps to provide a stringent constraint on the total amount of neutrino mass. As a recent indication~\cite{T2K:2019bcf}, a scenario in which leptonic CP symmetry is violated, is exciting since this may hold an essential key to understanding the matter-antimatter imbalance in the Universe.  Massive neutrinos also impact on the astrophysical processes such as supernova and the evolution of the stars and black holes. In short, neutrinos are a unique window into the Universe at its two extremes\textemdash the smallest and the largest.    
\section{Neutrino oscillation and measurements }
Our understanding of neutrino mass and leptonic mixing mainly comes from the neutrino oscillation measurements. Detailed description of neutrino oscillation phenomenology can be found elsewhere~\cite{ParticleDataGroup:2022pth}.  Neutrino oscillations require the existence of a neutrino mass spectrum, i.e mass eigenstate $\nu_i$ with definite mass $m_i$ where $i$ is 1,2, or 3. It requires flavor eigenstate with definite flavor, $\nu_{\alpha}$ where $\alpha$ is $e,\ \mu, \ \tau$ must be a linear superposition of the mass eigenstates 
$|\nu_{\alpha}\rangle = \sum_{i}U^*_{\alpha i}|\nu_{i}\rangle$ where $U_{\alpha i}$ is an unitary matrix characterizing the leptonic mixing, called the PMNS matrix. The probabilities of flavor transitions, $P(\nu_{\alpha} \rightarrow \nu_{\beta})=f\left(U_{\alpha i},U_{\beta i},\sin (\Delta m^2_{ij}\frac{L}{4E})\right)$, can be derived using the quantum mechanics. 
In matter, neutrino oscillation probabilities depend on the matter density since the scattering of electron (anti-) neutrinos with electrons in the matter induces an extra interaction potential energy into the Hamiltonian of the system. The matter effect is more pronounced with higher energy, longer baseline, and higher matter density. It is important to stress that the matter effect is not the same for neutrinos and anti-neutrino,  mimicking the \emph{genuine} CP violation effect, and being sensitive to neutrino mass ordering. The leptonic mixing matrix $U_{\alpha i}$ is a $3\times3$ unitary matrix parameterized with three mixing angles $(\theta_{12},\ \theta_{13},\ \theta_{23})$, one irreducible Dirac CP-violation phase \dcp, and two Majorana phases $(\rho_1,\ \rho_2)$  if neutrino is Majorana particle. 
The oscillation probabilities do not depend the $(\rho_1,\ \rho_2)$ so neutrino oscillation measurement is insensitive to these parameters as well as nature of neutrino mass. The experimental data~\cite{ParticleDataGroup:2022pth} have established that $\theta_{12} \approx \pi/6$; $\theta_{13} \approx \pi/20$, and $\theta_{23} \approx \pi/4$ while \dcp\ is not solidly established to be non-zero. The oscillated pattern is graphically described via the $\sin^2 \left(\Delta m^2_{ij}\frac{L}{4E}\right)$ term, whether $\Delta m^2_{ij}\frac{L}{4E} \equiv 1.27 \frac{\Delta m^2_{ij}[\text{eV}^2].L[\text{km}]}{E[\text{GeV}]}$ is used for convenience. The oscillation wavelengths, defined as $L_{\text{osc.}}[\text{km}]\equiv \frac{\pi}{2}\frac{E[\text{GeV}]}{1.27\Delta m^2_{ij}[\text{eV}^2]}$, are driven by two mass-squared splittings of the neutrinos
\begin{align}
\Delta m^2_{21} = m^2_2 - m^2_1 &\approx 7.4\times 10^{-5} (eV^2/c^4),\nonumber \\
|\Delta m^2_{31}| = |m^2_3-m^2_1|&\approx 2.5\times 10^{-3} (eV^2/c^4).
\end{align}
While both $\Delta m^2_{21},\ |\Delta m^2_{31}|$ are known with better than 3$\%$ precision, whether the mass ordering (MO) is $m_3>m_2>m_1$ (\emph{normal} MO) or $ m_2>m_1>m_3$ (\emph{inverted} MO) is still unknown with the present data. 

The neutrino oscillation measurements typically count the number $N^{\nu_{\beta}}$ of the $\beta$-flavor neutrino events as a function of reconstructed energy $E_{\nu}^{reco.}$ collected in a detector placed a $L$ distance from the neutrino source with average matter density of $\rho_{mat.}$. The counted event number is essentially proportional to the $\alpha$-flavor neutrino flux $\Phi_{flux}^{\nu_{\alpha}}$, the cross section $\epsilon_{det.}^{\nu_{\beta}}$, the detector mass $M_{det.}$, the detection efficiency $\epsilon_{det.}^{\nu_{\beta}}$, the oscillation probability $P(\nu_{\alpha}\rightarrow \nu_{\beta})$, and the detector response to the neutrino energy $R_{det.}(E_{\nu}^{true.},E_{\nu}^{reco.})$
\begin{align}\label{eq:eventrate}
N^{\nu_{\beta}}(E_{\nu}^{reco.},\vec{o}) = &\Phi_{flux}^{\nu_{\alpha}}(E_{\nu}^{true})\times \sigma_{int.}^{\nu_{\beta}}(E_{\nu}^{true})\times M_{det.} \times \epsilon_{det.}^{\nu_{\beta}} (E_{\nu}^{true}) \times R_{det.}(E_{\nu}^{true.}, E_{\nu}^{reco.}) \nonumber \\
&\times P(\nu_{\alpha}\rightarrow \nu_{\beta}|E_{\nu}^{true},\vec{o}),
\end{align}
where parameter vector $\vec{o} =(\Delta m^2_{21}, \Delta m^2_{31};\theta_{12},\theta_{13},\theta_{23},\delta_{CP};\rho_{mat.}, L)$. From Eq.~\ref{eq:eventrate}, it becomes clear that for a precise measurement of neutrino oscillations we need (i)powerful and well-controlled sources of (anti-)neutrinos, (ii) big detectors with flavor-tagging and energy-reconstruction capabilities, (iii) well-modeled $\nu/\bar{\nu}$ interactions with nucleons and detector's  $\nu/\bar{\nu}$  energy resolution, and (iv) capability to resolve the parameter degeneracies, particularly among \dcp,\ \threactor, \thatm, $sign(\Delta m^2_{31})$. The ability to resolve parameter degeneracy is critical for precision measurement because no single experiment can be sensitive to all parameters and oscillation parameters are extracted from the measurement of oscillation probabilities, which is a function of multiple parameters, resulting in multiple clone solutions. In the T2K and NOvA experiments, sources of muon (anti-)neutrinos are used for study and the flavors, which can tagged efficiently in the far detector, are muon and electron. Thus, both experiments can measure four main transitions $\nu_{\mu}\rightarrow \nu_{\mu}$  and its CPT-mirror image $\overline{\nu}_{\mu}\rightarrow \overline{\nu}_{\mu}$,  $\nu_{\mu}\rightarrow \nu_{e}$ and its CP-mirror image $\overline{\nu}_{\mu}\rightarrow \overline{\nu}_{e}$.

Neutrino oscillation measurements have yield accurate results for three mixing and two mass-squared splittings. However, the picture of leptonic mixing and neutrino mass spectrum is not yet complete. We do not know yet what the neutrino mass ordering is, whether CP symmetry is violated in the lepton sector, how close the leptonic mixing angle \thatm\ is to $\pi/4$; where the sterile neutrino exists, and whereas the 3x3 unitary mixing matrix is an exact or effective framework at the low-energy scale. The on-going and future neutrino oscillation experiments will shed the light on these unknowns. Other puzzles, like as absolute mass of neutrinos and the origin of neutrino mass,  require different types of experiments, which are beyond the scope of this session.  

\section{T2K and NOvA experiments}

\noindent \emph{\textemdash Experimental specifications}

\begin{figure}[H]
\centering
\includegraphics[width=0.8\textwidth]{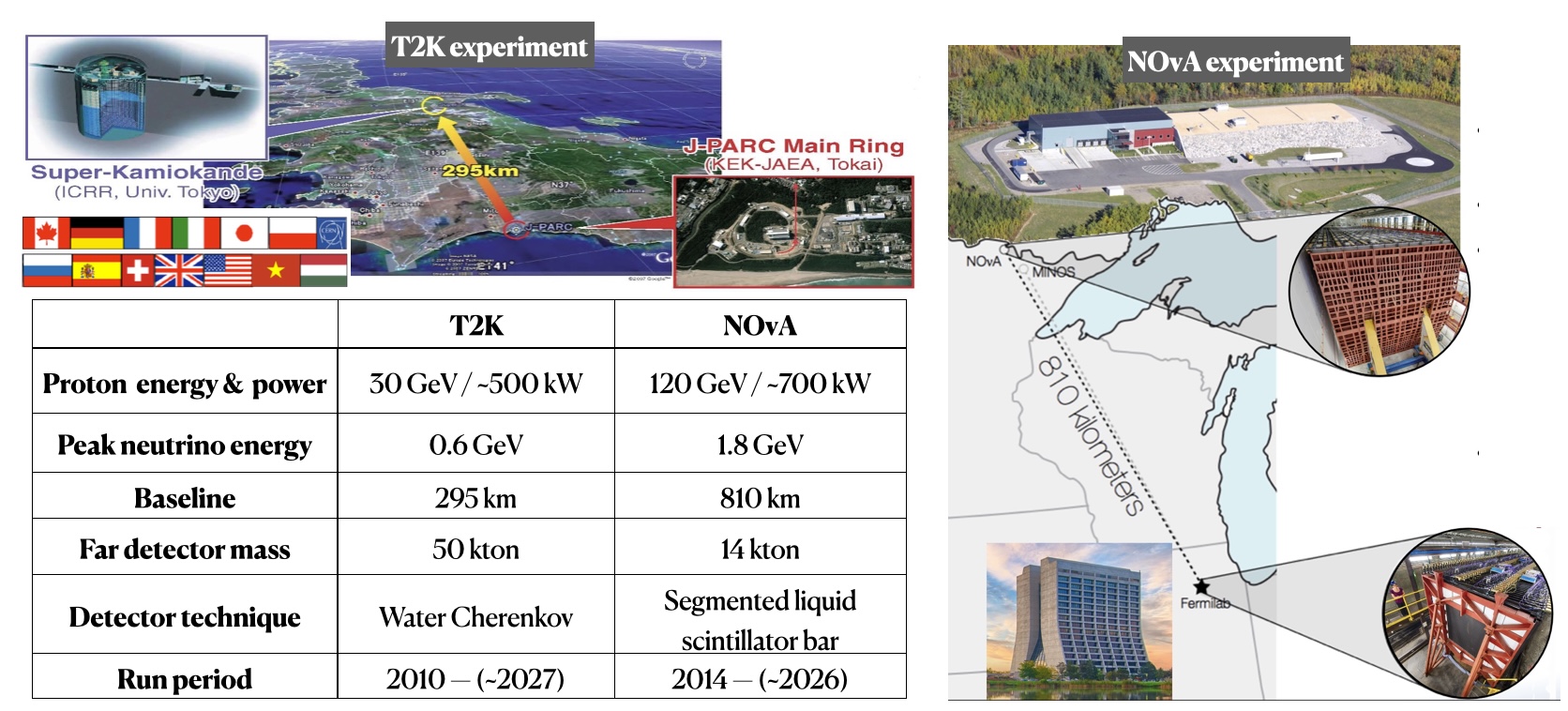} 
\caption[]{Experimental layout and main specifications of the T2K and NOvA experiments}
\label{fig:t2knovaexp}
\end{figure}

Details of T2K and NOvA experiments~\cite{Abe:2011ks,NOvA:2007rmc} can be found elsewhere. The two experiments continue their predecessors K2K~\cite{K2K:2002icj} and MINOS(+)~\cite{MINOS:2008hdf} and play important roles in the development of the third generation of neutrino experiments Hyper-Kamiokande~\cite{Hyper-Kamiokande:2018ofw} and DUNE~\cite{DUNE:2015lol}. T2K and NOvA experiments share a similar concept of neutrino oscillation measurement: a highly pure and intense beam of (anti-)muon neutrinos, produced by a powerful beam of accelerated high-energy protons on a fixed target, is measured at near-site and far-site detectors to uncover the oscillation pattern and extract the underlying physics parameters. T2K (NOVA) far detector is placed 295~km (810~km) from the neutrino production source respectively. These experimental baselines are chosen to be close to the oscillation wavelength $L_{\text{osc.}}$ driven by the atmospheric mass-squared splitting $\Delta m^2_{32}$ for a neutrino flux with the energy peak of 0.6~GeV (1.8~GeV) for T2K (NOvA) respectively. Both take advantage of using the near detector complex to understand better the neutrino source composition and neutrino interactions, which is essential for unraveling the neutrino oscillation effect from the others at the far detectors. The experimental layout and main specifications are shown in Fig.~\ref{fig:t2knovaexp}. 

 The two experiments use different techniques for detecting neutrino interactions and tagging the involved flavor of neutrinos.  T2K ultilizes 50~kton water Cherenkov detector, Super-Kamiokande~\cite{Super-Kamiokande:2002weg}, while the NOvA uses a 14~kton far detector made of segmented liquid scintillator PVC-supported tubes.  T2K began operation in 2010 and is expected to continue until roundly 2027, whilst NOvA started collecting physics data in 2014 and is expected to continue until 2026. Complementary in neutrino oscillation measurements, particularly for the CP violation search and the neutrino mass ordering determination, between T2K and NOvA experiments can be understood using an approximate formula of the asymmetry in the probabilities of $\nu_{\mu}\rightarrow \nu_e$ transition and its CP-mirror image $\overline{\nu}_{\mu}\rightarrow \overline{\nu}_e$
 \begin{align} \label{eq:adcp} 
    A_{\text{CP}} \equiv \frac{P_{(\nu_{\mu}\rightarrow \nu_e)}- P_{(\bar{\nu}_{\mu}\rightarrow \bar{\nu}_e)}}{P_{(\nu_{\mu}\rightarrow \nu_e)} + P_{(\bar{\nu}_{\mu}\rightarrow \bar{\nu}_e)}} 
    &\approx -0.256 \sin \delta_{\text{CP}} \pm \frac{L}{\text{2800km}},
\end{align}
where $\pm$ is for \emph{normal} (\emph{inverted}) neutrino mass ordering respectively. The first term in Eq.~\ref{eq:adcp}, induced by \dcp\ values, presents the \emph{genuine} CP violation effect, whereas the second, induced by the matter effect, is regarded as \emph{fake} CP violation impact. For T2K and NOvA, the \emph{genuine} CP asymmetry, which are proportional to $\sin \delta_{\text{CP}}$, are at similar order and about 25.6$\%$. According to Eq.~\ref{eq:adcp}, the \emph{fake} CP asymmetry presents in the 295km-baseline T2K (810km-baseline NOvA) is  10.5\% (28.9\%) respectively. A joint analysis between T2K and NOvA is expected to enhance the search for CP, resolve the ambiguity in the neutrino mass ordering, and provide a more complete picture of the leptonic mixing. Furthermore, data from the two experiments are valuable for understanding the models of accelerator-based neutrino flux and neutrino-nucleon(nucleus) interactions. 

\noindent \emph{\textemdash Accelerator-based neutrino beams}

\begin{figure}[H]
\centering
\includegraphics[width=0.8\textwidth]{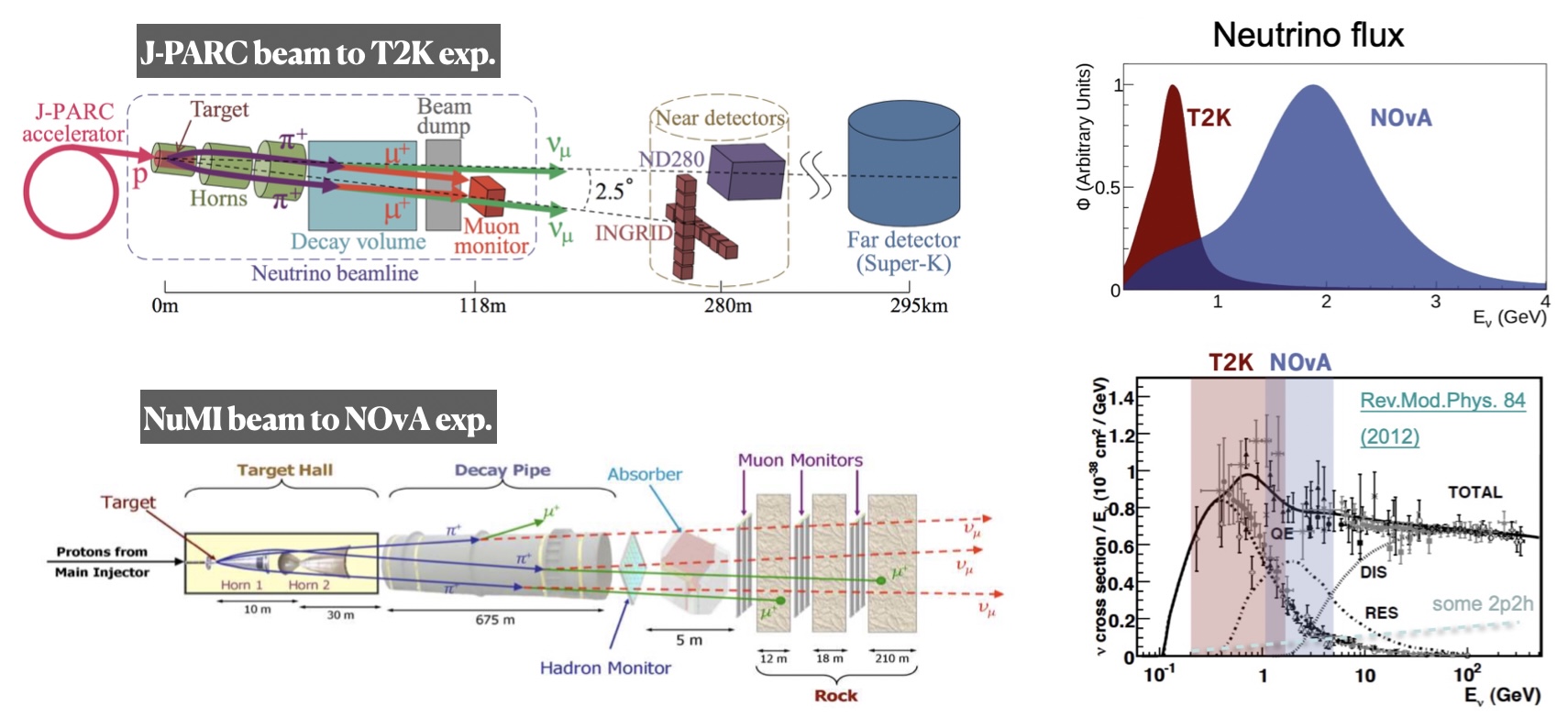} 
\caption[]{Layout of the J-PARC neutrino beam and NuMI beam to T2K and NOvA detector respectively. The off-axis technique leads to the narrow band of the produced neutrino.}
\label{fig:beam}
\end{figure}

To create the neutrino sources for T2K (NOvA), high energy protons of 30~GeV (120~GeV) from the accelerator are extracted, guided, and bombarded onto a target of (91.4~cm (94.0~cm)-length graphite rod respectively. The produced charged hadrons (mainly $\pi,\ K$) are focused by a magnetic horn system and decayed in flight into neutrinos and secondary hardrons. A hadron absorber or beam dump is placed downstream of the decay pipe ( approximately 100~m (675~m) for T2K (NOvA)) to filter out the hardrons and deliver highly pure neutrino beam to the experiment's detectors. The layout of T2K and NOvA neutrino beam is shown in Fig.~\ref{fig:beam}.  One important feature of the magnetic horn systems is flexibility to switch their polarity, consequently focusing positive or negative hadrons at wish to produce mainly \numu\ or \numubar\  beam. The feature allows T2K and NOvA test both CPT invariance and leptonic CP invariance in a single experiment.  Both T2K and NOvA far detectors are placed 2.5~mrad and 14.6~mrad off-axis to harness a narrow-band neutrino beam\textemdash a handy technique to enhance the signal-to-noise ratio in the observation and substantially improve the precision of the measurement. 

\noindent To predict the neutrino flux and its uncertainty, a dedicated simulation program has been developed to take into account the real-time condition of the primary proton beam, the data-driven model of hadron production from proton-target interaction, and the transport of particles through the beamline component. Detail of neutrino flux and its recent improvement can be found in Ref.~\cite{T2K:2023smv,NOvA:2021nfi}. To sum up, T2K and NOvA experiments can deliver a highly pure ($>$90\%) \numu\ and  \numubar\ beam with few percents of \emph{wrong-sign} components (eg. \numubar\ in the neutrino beam and \numu\ in the anti-neutrino beam) and less than 1\% of intrinsic \nue\ / \nuebar\ components. Uncertainty of the flux prediction on the event rate at far detector is about 6-8\% and further suppressed thank to the constraint from the near detector measurement. The dominant uncertainty is the hadron production and further improvement is expected in the near future. 

\noindent \emph{\textemdash Neutrino-nucleus interaction}

\noindent Understanding the neutrino interaction model is critical for measuring neutrino oscillations since the flavor and energy of entering neutrinos are inferred only from the observable product of neutrino-nucleon(nuclei) interaction inside the detector. The variety of neutrino interaction types, the lack of understanding of the nuclear medium effect, and imperfection in the detector response complicate this inference. In T2K, the dominated interaction type is the charged-current quasi-elastic (CCQE) interaction and considerable contributions from the multi-nucleon knockout or two-particle two-hole (2p2h) interaction and the CC resonance production. With higher neutrino energy range, NOvA has large contribution from the CC resonance and deep-inelastic scattering (DIS). Sophisticated neutrino-nucleus interaction event generators are used by the experiment to implement the theoretical models for individual neutrino interaction types, to take into account the effect of nuclear medium on the initial state of primary neutrino-nucleon interaction, and to produce at event level the list of final-state particles which can be tracked by the detector Monte Carlo (MC) simulation. The uncertainty of neutrino interaction model, which is typically convoluted with uncertainty in the neutrino flux due to energy-dependence, results in significant uncertainty on the total event rate (eg. roughly 13\% in T2K) and consequently affect the precision of oscillation parameter measurement. A practical technique to mitigating the impact is to make use of the near detector's high-statistic data samples to tune and constrain the neutrino interaction model (\emph{and indispensable neutrino flux}) before extrapolating to what expected to be observed at the far detector. The two-detector technique effectively reduces, for example in case of T2K experiment, from 13\% to 3-6\% (depend on the data sample) uncertainty of a combined neutrino interaction model and flux on the total event rate.  T2K and NOvA use notably different interaction models~\cite{T2K:2023smv,NOvA:2021nfi} with independent neutrino event generators as nominal MC simulation and the consistency and robustness must be cross-checked conscientiously for a joint data analysis.

\noindent \emph{\textemdash Event reconstruction, selection, and data samples}

The charged particles induced from the neutrino-nucleus interaction are tracked and reconstructed in order to provide essential information of the incident neutrinos. Charged particle with sufficient momentum in the T2K far detector typically forms a light ring pattern seen by the single-photon resolving 50~cm-diameter vacuum-based PMT. The observed ring pattern can differentiate muon and electron due to their difference in electromagnetic activities. Basically the electron ring is fuzzier than the muon ring. In addition, T2K far detector can identify electron from muon decay by finding the late time-clustered hits in event-by-event basic with high efficiency (roughly 96\% for $\mu^{+}$ and 86\% for $\mu^{-}$.) This feature allows T2K to distinguish CCQE neutrino interactions, in which neutrino energy can be reconstructed precisely from the kinematics of the outgoing lepton, from the other. In the presented analysis, T2K selects five data samples, including four CCQE-like samples of \numu\ and \nue\ in the neutrino and anti-neutrino beams, and one CC single pion-like \nue\ in the neutrino beam, detailed in Ref.~\cite{T2K:2023smv}. NOvA near and far detectors function as the tracking calorimeters with mineral oil-based liquid scintillator, wavelength shifting fiber, and multi-pixel avalanche photodiode operated at low temperature.  Essentially, muon-like track is longer and less scattering than the electron-like track. Since NOvA's far detector is placed on the surface with modest overburden, it is necessary to preselect the data sample by filtering out 130kHz cosmic ray using beam timing and direction. NOvA experiment adopts strongly the machine and deep learning to classify the neutrino flavors and interactions. Muon neutrino-like energy is reconstructed by summing of the track length-based muon energy and total calorimetric energy of produced hadrons while electron neutrino-like energy is a quadratic function of identified electromagnetic shower and remaining hadronic activities. Four main data samples, consist of \numu\ and \nue-like events in neutrino and anti-neutrino beams are further categorized into multiple sub-sample using the quantitative measures of particle identity and hadronic activity to enhance the sensitivity and precision and the involved oscillation parameters,  detailed in Ref.~\cite{NOvA:2021nfi}.

\section{Results }
In this report, the result is based on T2K data samples collected from delivery of a total $36\times 10^{20}$ protons-on-target (POT), which includes $19.7\times 10^{20}$ POT and $16.3\times 10^{20}$ POT taken with neutrino and anti-neutrino beam respectively. For NOvA, the analyses use $13.6\times 10^{20}$ POT in neutrino beam and $12.5\times 10^{20}$ in anti-neutrino beam. Most results present in this proceedings can be found in Ref.~\cite{T2K:2023smv,T2K:2023mcm,NOvA:2021nfi}. Here we highlight the main results. 

\noindent \emph{\textemdash Universality of the standard 3-flavor framework}

\begin{figure}[H]
\centering
\includegraphics[width=0.4\textwidth]{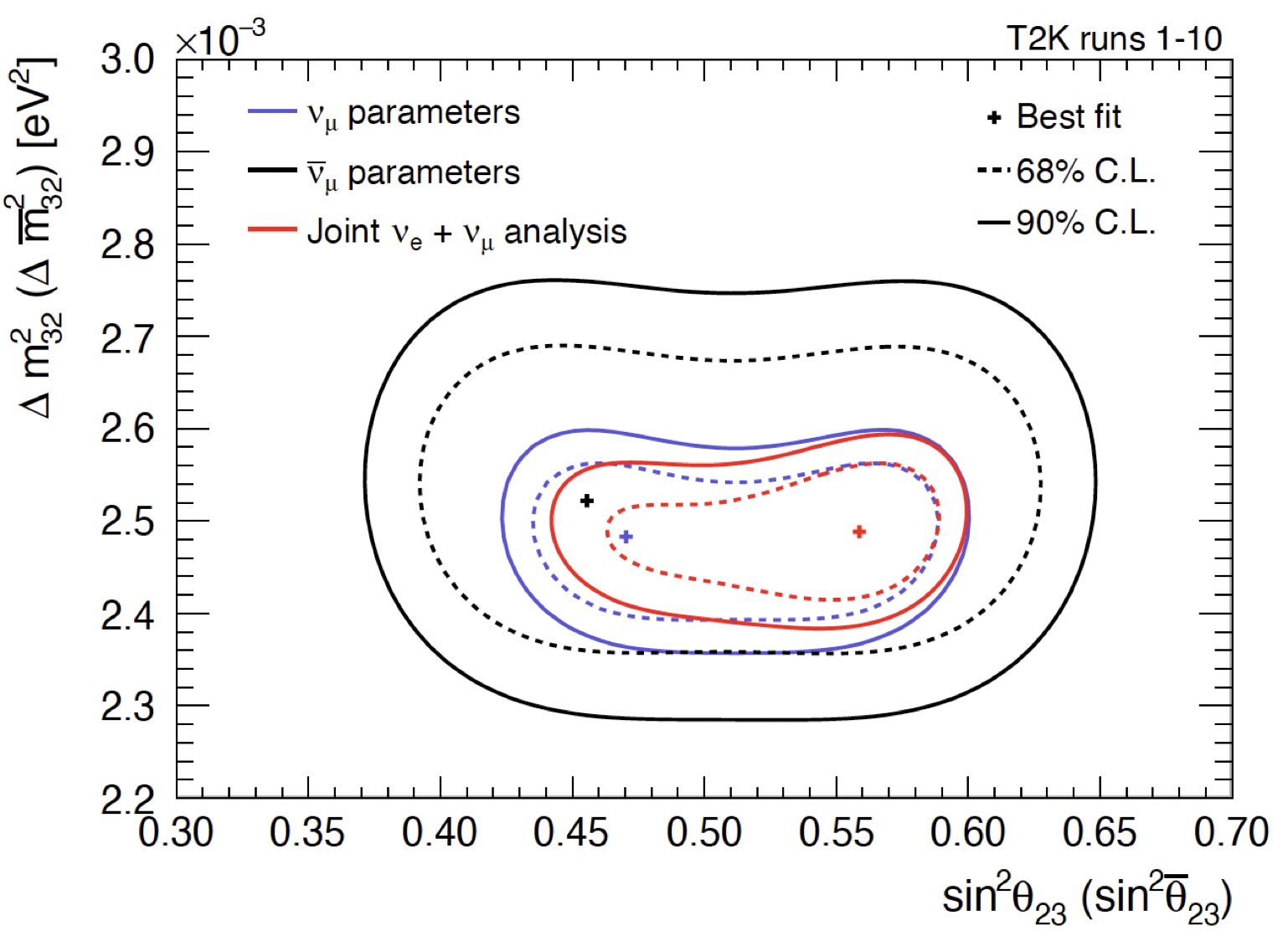} 
\includegraphics[width=0.4\textwidth]{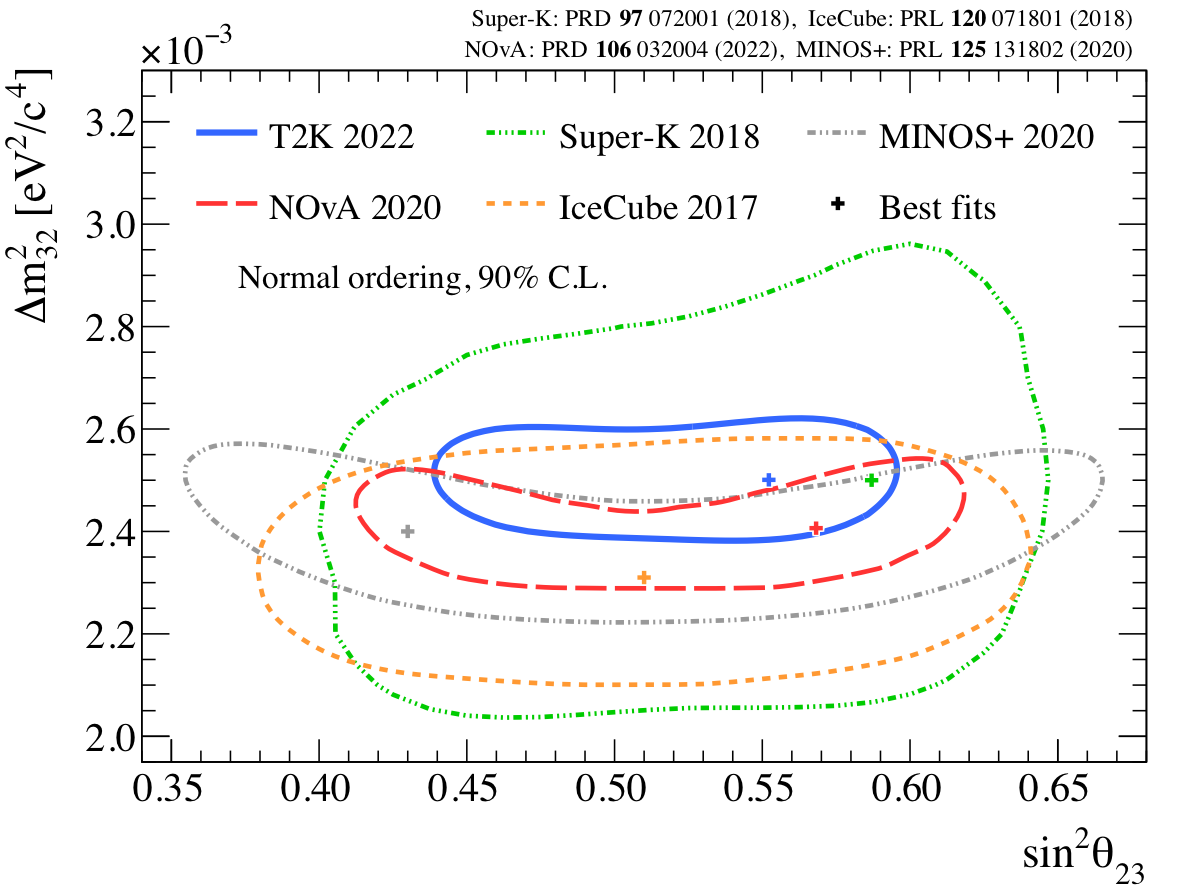} 
\caption[]{The left shows T2K results on the disappearance of muon-neutrino and muon-antineutrino assuming an independent set of parameters governing the underlying oscillation pattern. The right shows the consistence between T2K, NOvA and other experiments, taken from Ref.~\cite{T2K:2023mcm}}
\label{fig:t2kdis}
\end{figure}

\noindent Firstly, we present results from measuring the disappearance of \numu\ and \numubar\ from the beams of almost pure \numu\ and \numubar\ respectively. The measurements are motivated by the CPT invariance test of the neutrino and anti-neutrino mass spectra. In this analysis, the oscillation parameter sets, which drive the \numu\ and \numubar\ disappearance, are treated independently. Fig.~\ref{fig:t2kdis} shows that these two parameter sets for neutrinos and antineutrinos measured by T2K agree with each other, implying a consistency with the CPT invariance hypothesis. 
\noindent Particularly, the best-fit values for neutrinos are $(\sin^2\theta_{23}=0.47^{+0.11}_{-0.02}, \Delta m^2_{32}=2.48^{+0.05}_{-0.06}\times 10^{-3}\text{eV}^2)$, whilst those for anti-neutrinos are $(\sin^2\overline{\theta}_{23}=0.45^{+0.16}_{-0.04}, \Delta \overline{m}^2_{32}=2.53^{+0.10}_{-0.11}\times 10^{-3}\text{eV}^2)$. Assuming that the CPT is hold, the neutrino and anti-neutrino data can be combined to provide the most constraint on these atmospheric parameter. For T2K, most precise values are $(\sin^2\theta_{23}=0.467^{+0.106}_{-0.018}, \Delta m^2_{32}=2.495^{+0.041}_{-0.0658}\times 10^{-3}\text{eV}^2)$, driven by the neutrino data. For NOvA, the best-fit values for the atmospheric parameter is. $(\sin^2\theta_{23}=0.57^{+0.03}_{-0.04}, \Delta m^2_{32}=2.41\pm 0.07\times 10^{-3}\text{eV}^2)$. Fig.~\ref{fig:t2kdis} shows a consistency in the parameter \sinsqthatm-$\Delta m^2_{32}$ among various experiments.
\noindent Besides, no statistically significant indication of non-standard physics, such as existence of sterile neutrinos, heavy neutral particle, or Lorentz invariance, is found in both T2K and NOvA experiments. To sum up, both neutrino and anti-neutrino data samples in both T2K and NOvA experiments can be well-fitted within a \emph{standard} 3x3 unitary mixing framework. 

\noindent \emph{\textemdash CP-violation, mass ordering, and $\theta_{23}$ mixing angle}

\begin{figure}[H]
\centering
\includegraphics[width=0.75\textwidth]{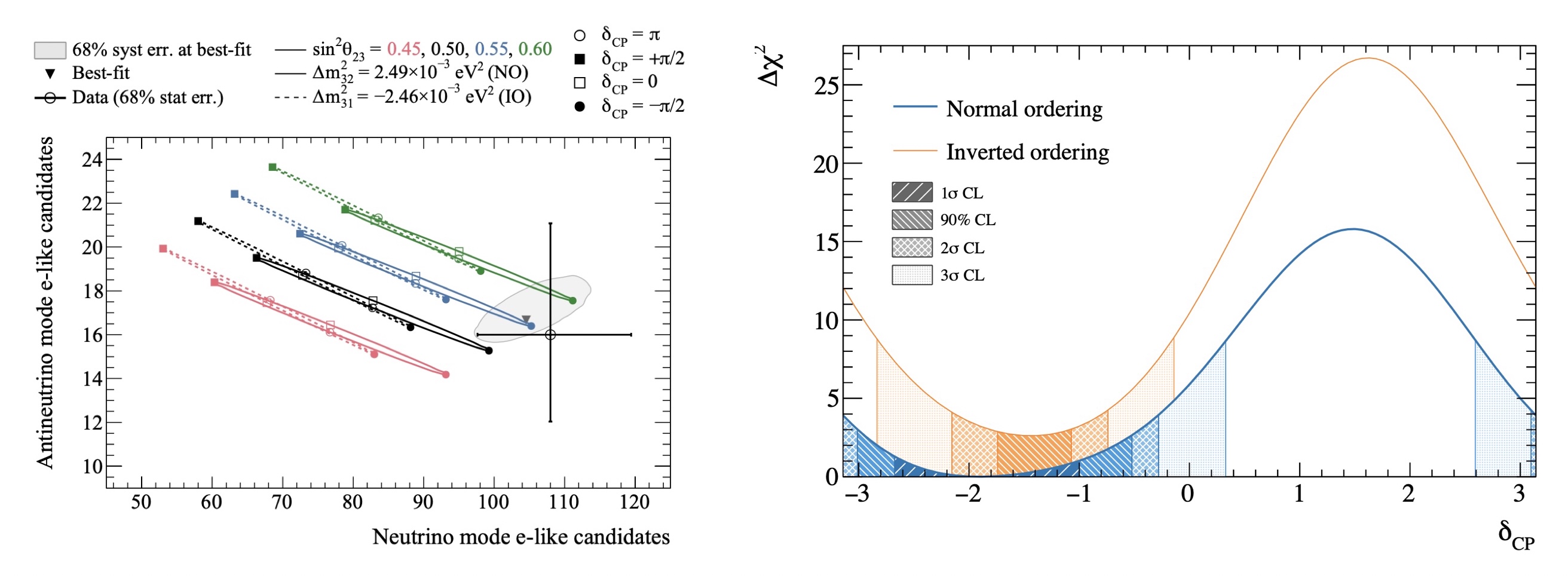}\\ 
\includegraphics[width=0.75\textwidth]{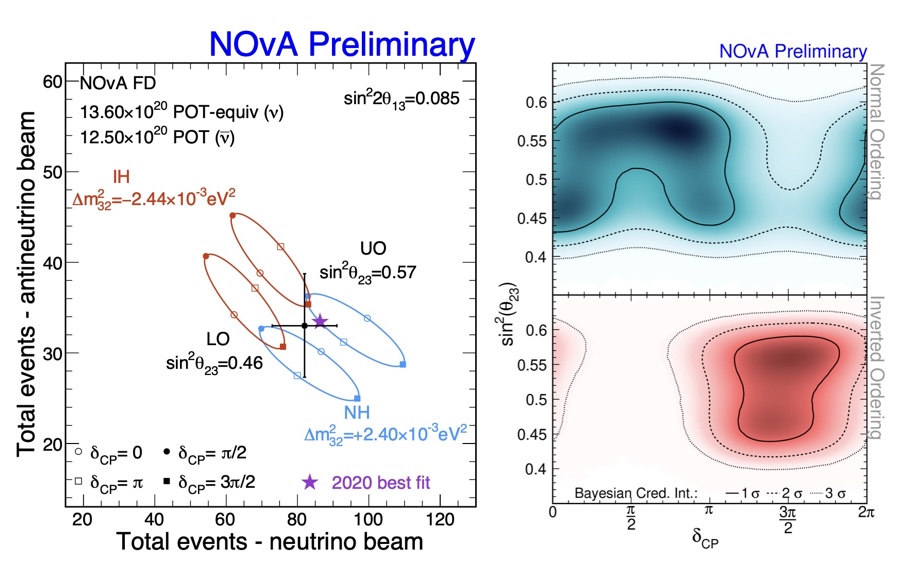} 
\caption[]{The top left shows T2K's $\nu_e (\overline{\nu}_e)$ appearance data samples from the neutrino (anti-neutrino) beam  and the expected events at different combined set of neutrino mass ordering, \dcp, and \sinsqthatm. The top right plot shows the T2K's allowed region of \dcp\ at different C.L. for \emph{normal} (\emph{inverted}) MO. The corresponding  data and analysis results from NOvA are showed in the bottom plots.}
\label{fig:t2kdcp}
\end{figure}

\noindent The leptonic CP violation, manifests itself as the \dcp\ irreducible phase, can be measured by comparing the probabilities of two CP-mirrored processes, practically $\nu_{\mu}\rightarrow\nu_e$ and $\overline{\nu}_{\mu}\rightarrow\overline{\nu}_e$. The relative difference between the two process probabilities, as shown in Eq.~\ref{eq:adcp}, is proportional to $\sin\delta_{CP}$ and depends on the neutrino mass ordering. Besides, a closer look at the analytical formula of the  $\nu_{\mu}\rightarrow\nu_e$ and $\overline{\nu}_{\mu}\rightarrow\overline{\nu}_e$ probabilities exhibit a non-trivial correlation of \dcp with \sinsqthatm\ and \sinsqthreactor, which has a strong impact on the CP violation study. Another major challenging is statistics, because the process probabilities are typically at few percent level.  Fig.~\ref{fig:t2kdcp} depicts the an event rate comparison in T2K analysis between data and MC prediction at different values of \dcp, mass ordering, and \sinsqthatm. T2K observed 94 (16) CCQE-like candidates of $\nu_{\mu}\rightarrow \nu_e$ ( $\overline{\nu}_{\mu}\rightarrow \overline{\nu}_e$) transitions, which significantly differ from prediction at CP-conserving values of \dcp\ (for reference 81.59 (18.81) at \dcp=0 and 81.85  (18.49) at \dcp=$\pi$ with systematic uncertainties of 4.7\% (5.9\%)). A fine analysis results in a statistical test $\Delta \chi^2$ profile as function of \dcp with two hypothesis of neutrino mass ordering, as showed in Fig.~\ref{fig:t2kdcp}.
A large fraction of \dcp\ around $\pi/2$ is excluded at 3$\sigma$ C.L. or higher. The CP-conservation values (0,$\pi$) of \dcp\ is excluded at 90\% C.L. but still be allowed in 2$\sigma$. Another measure of CP-violation amplitude is via Jarlskog invariance, $
J_{\text{CP}}^{\text{Lepton}}= \frac{1}{8} \sin2\theta_{12}\sin2\theta_{13}\cos\theta_{13}\sin2\theta_{23}\sin\delta_{CP}
$, which does not depend on the matrix parameterization. T2K data favor a negative values of $J_{\text{CP}}^{\text{Lepton}}$ at $\approx$-0.033. Comparing to the Jarlkskog invariance in the quark mixing~\cite{ParticleDataGroup:2022pth} $J_{\text{CP}}^{\text{Quark}}=(3.08^{+0.15}_{-0.13})\times 10^{-5}$, the $J_{\text{CP}}^{\text{Lepton}}$ can be two order larger, which is interesting since the matter dominance in the Universe can be explained entirely via the leptogenesis and see-saw mechanism.

\noindent NOvA's data and analytical result on the similar measurement is shown in bottom plots of Fig.~\ref{fig:t2kdcp}. NOvA observes 82 (33) candidates of $\nu_{\mu}\rightarrow \nu_e$ ( $\overline{\nu}_{\mu}\rightarrow \overline{\nu}_e$) transitions, which does not display a significant deviation from the prediction with CP-conservation hypothesis. 
\noindent Regarding to the neutrino mass ordering, both T2K and NOvA weakly prefer the \emph{normal} over \emph{inverted} hypotheses. Also, the higher octant of \sinsqthatm\ is mildly favored. The results are summarized in Table~\ref{tab:moth23}.
\begin{table}[H]
\begin{center}
  \begin{tabular}{|l|l|l|l|l|}
    \hline
    \multirow{2}{*}{Mass ordering} &
      \multicolumn{2}{c|}{T2K} &
      \multicolumn{2}{c|}{NOvA} \\ \cline{2-5}
 & \sinsqthatm$<0.5$ & \sinsqthatm$>0.5$ & \sinsqthatm$<0.5$ & \sinsqthatm$>0.5$  \\
    \hline
    \emph{Normal} & 20\% & 54\% & 26\% & 42\%  \\
    \hline
    \emph{Inverted} & 15\% & 21\% & 11\% & 21\%  \\
    \hline
  \end{tabular}
  \caption{Fraction of T2K and NOVA posterior probability integrated for four combined (MO, \thatm\ octant) sets.}
  \label{tab:moth23}
  \end{center}
\end{table}


\section{Summary and prospects}
We have presented the results of the neutrino oscillation measurements from T2K and NOvA experiments, based on data and analyses completed by summer 2023. The experimental data can be described well in the \emph{standard} 3x3 unitary mixing framework. T2K provides a significant hint on the CP violation, whereas NOvA shows dissimilar tendency if neutrino mass ordering is \emph{normal}. If neutrino mass ordering is \emph{inverted}, the two experiments consistently favor the maximal CP violation. \emph{Normal} mass ordering and higher octant of the \thatm\ mixing angle are weakly preferred for both  experiments.  

T2K and NOvA have been collecting more data than the given results and are now reviewing it in order to provide updated measurements. Both experiments have worked hard to improve neutrino beam power, detector performance, and neutrino interaction models. T2K and NOvA have collaborated on a combined analysis, and the results are expected to be released soon. T2K and Super-Kamiokande have also begun collaboration on a joint study of accelerator-based and atmospheric neutrino sources in a single detector. These collaborative studies are expected to yield unprecedented sensitivity to leptonic CP violation, neutrino mass ordering, and other neutrino oscillation parameters, as well as a thorough investigation of the \emph{standard} 3x3 unitary mixing paradigm. 


\end{document}